\documentclass[preprint,12pt,authoryear,harvard]{elsarticle}
\usepackage{graphicx}  
\usepackage{dcolumn}   
\usepackage{bm}        
\usepackage{amssymb}   

\usepackage[]{natbib}
\journal{Physica E}

\begin{document}

\newcommand{\etal}{{\em et al.}}{}
\newcommand{\eqn}[1]{Eq.~\ref{#1}}
\newcommand{\fig}[1]{Fig.~\ref{#1}}
\newcommand{\tab}[1]{{Table ~\ref{#1}}}
\newcommand{\AN}{$\rm{\AA}$\,}{}
\newcommand{\ww}{0.99}
\newcommand{\wT}{0.19}
\newcommand{\wh}{0.27}
\newcommand{\paper}{paper\,}

\begin{frontmatter}

\title{Tunable Band Gaps of Mono-layer Hexagonal BNC Heterostructures}

\author[rvt]{Qing Peng\corref{cor1}\fnref{fn1}}
\ead{qpeng.org@gmail.com}
\ead[url]{http://qpeng.org}
\fntext[fn1]{Tel: 1 (518) 279-6669 Fax: 1 (518) 276-6025 Mail: 110 8th Street JEC 2303, Troy, NY 12180 }
\author{Suvranu De}
\cortext[cor1]{Corresponding author}

\address{
 Department of Mechanical, Aerospace and Nuclear Engineering, Rensselaer Polytechnic Institute, Troy, NY 12180, U.S.A. \\
}

\begin{abstract}
We present an {\em ab initio} density functional theory (DFT)-based study of 
h-BN domain size effect on band gap of mono-layer h-BNC heterostructure modeled as (B$_3$N$_3$)$_x$(C$_6$)$_{1-x}$.
The atomic structures, electronic band structures, density of states 
and electron localization functions of h-BNC 
are examined as h-BN concentration ranged from 0 to 100\%.  
We report that the electronic band gap energy of h-BNC can be continuously 
tuned in full range between that of two phases, 
graphene and h-BN, as a function of h-BN concentration.
The origin of the tunable band gap in these heterosturcutes 
are due to the change in the electron localization with h-BN concentration. 
\end{abstract}


\end{frontmatter}


\section{INTRODUCTION}
Many approaches have been proposed to improve the semiconducting properties of graphene including the use of 
strains \citep{PhysRevB.78.075435},
patterned defects \citep{PhysRevB.82.073410},
dimensionality reduction as nanoribbons 
external supper-lattice potentials\citep{PhysReVB.79.205435}, 
doping and(or) application of external electric fields \citep{Ohta2006},
addition of chemical species such as hydrogen \citep{ISI:000262862800036,ISI:000265030000048}, 
substrate surface chemistry \citep{ISI:000262724000014},
and introduction of antidote lattice structures \citep{PhysReVLett.100.136804}. 
Among all these methods, one particularly promising technique has been reported to be the doping of B and N atoms into the graphene lattice. 
Experimental and theoretical studies show that both p-type and n-type semi-conducting graphene 
may be generated by substituting the C atoms with B and N atoms, respectively
\citep{ISI:000265832400041}. 
In such methods the dopant B and N atoms serve to modify the electronic structure of the graphene and increase the band gap.  

Hexagonal boron nitride (h-BN) and graphene have similar 2D lattice structures but with very different physical properties. 
Graphene has a zero band gap while h-BN is an insulator with a wide band gap (5.5 eV in bulk \citep{Song:3209}). 
Interesting structure can be made by mixing these two structures
\citep{Suenaga24101997}. 
It has been suggested that inserting a graphene layer between layers of hexagonal BN (h-BN) can exhibit tunable band gap as large as 0.23 eV \citep{PhysRevB.82.085431}. 
A small band gap can be also opened by placing a graphene layer onto a h-BN substrate \citep{Giovannetti}.

Recently, a promising method has been reported in which atomic mono-layers have been generated 
consisting of h-BN phases in graphene (h-BNC) using a thermal catalytic chemical vapor deposition method \citep{ISI:000276953500024}. 
This hybrid mono-layer has been shown to have isotropic physical properties which can be tailored 
by controlling the kinetic factors affecting the h-BN domain size within graphene. 
This is different from B-doped or N-doped graphene, where the integrity of the h-BN structure is missing. 
Although there are a few studies about domain-wise phase segregation in such B-C-N graphene-like layers
\citep{ISI:000273204200013,ISI:000283222200071,ISI:000290914700066,ISI:000286487300049}, 
a systematic study of the electronic band gap with respect to the h-BN domain size is missing.

The method of Adding h-BN phase is much more efficient in opening band gap in
graphene than all other methods. There are two most
important advantages in this method: (1) wide range of
tunable band gap of up to 4.7 eV can be obtained, 
which is one order of
magnitude larger than other methods; (2) integrity of
the individual phases is kept which make it much
easier for fabrication. 
In this paper, we examine the h-BN domain size effect on band gap  
of h-BNC monolayer using {\em ab initio} density functional theory. 
The electronic band gap with respect to the h-BN domain concentrations is studied 
and the origin of the tunable band gap in these heterosturcutes is also discussed.

\section{MODEL} 

The h-BN domain in h-BNC hybrid structures is modeled by the concentrations of h-BN
while maintaining the hexagonal structure of BN locally within the system.
The h-BN domain size effect
then can be represented by the h-BN concentration $x$ in the
model as (B$_3$N$_3$)$_x$(C$_6$)$_{1-x}$ where (B$_3$N$_3$) and (C$_6$)
denote the nanodomain structure of h-BN monolayer and graphene, respectively.
The proposed domain size effect model (B$_3$N$_3$)$_x$(C$_6$)$_{1-x}$ is based on
the results of previous studies of B$_x$C$_y$N$_z$ layered structures\citep{Mazzoni2006},
layers and nanotubes\citep{ISI:000286487300049,ISI:000290652200008},
quantum dots and nanorods\citep{ISI:000265030000048}, and monolayer nanohybrids\citep{ISI:000290914700066}.
It is a general believe that there are domain-wise segregation phases of h-BN and graphene
with lower energy, larger band gap, and better thermal stability \citep{ISI:000276953500024,ISI:000286487300049,ISI:000290914700066}.
As a result, the six-atom hexagonal structures (both B$_3$N$_3$ and C$_6$)
are the basic blocks in these heterostructures.
This model of (B$_3$N$_3$)$_x$(C$_6$)$_{1-x}$ captured the main feature of these heterostructures.
It was found that both stoichiometry and geometry vary the band gaps of B-C-N nanotube and layered materials
and the maximum band gap is achieved at B/N ratio of 1 \citep{Mazzoni2006,ISI:000286487300049}.
while in B-C-N monolayer,
 domain size is a dominant factor comparing with domain shape in tuning band gaps \citep{ISI:000290914700066}.
By varying h-BN concentration $x$, the domain size effect
on mechanical properties of mono-layer hexagonal BNC heterostructures
can be appropriately
presented in our (B$_3$N$_3$)$_x$(C$_6$)$_{1-x}$ model.

We need to emphasis that although we used stoichiometry of h-BN here,
this domain model is different from point model of B$_x$C$_y$N$_z$ model
at $x=z$ where hexagonal (B$_3$N$_3$) structure is not considered, for example \citep{Mazzoni2006}.
In other words, our model specified not only the stoichiometry of B/N ration of 1,
but also the hexagonal (B$_3$N$_3$) and C$_6$ structures,
to represent the two separated phases in heterogeneous h-BNC structures.

We examined the variation of elastic band gaps of the $h$-BNC monolayer as function of $h$-BN concentration.
Five $h$-BNC configurations, in order of $h$-BN concentration, 0\%, 25\%, 50\%, 75\% and 100\% have been studied,
where 0\% and 100\% corresponds to the pure graphene and $h$-BN, respectively.
The three other concentrations were selected based on their simplicity and representativeness.
The atomic structures of these five configurations (\fig{fig:config})
were determined by the {\em ab initio} density functional theory.

Due to the segregation of two phases h-BN (B$_3$N$_3$) and graphene (C$_6$),
the minority will form ``island", as observed in experiments \citep{ISI:000276953500024}.
For a certain stoichiometry, as presented by h-BN concentration $x$,
the larger system size used in the simulation box,
the larger ``island" will be formed to achieve lower energy and become more stable.
In this study, we used a small system of 24 atoms due to its simplicity while containing
its distinguish characteristics of modeling separated phases in h-BNC heterostructures.
The system size effect of this model was examined
in a larger system with 150 atoms.

\section{COMPUTATIONAL DETAILS} 

DFT calculations were carried out with the Vienna Ab-initio
Simulation Package (VASP)
\citep{VASPa} which is based
on the Kohn-Sham Density Functional Theory (KS-DFT) \citep{DFTa} with the generalized gradient approximations as parameterized by Perdew, Burke and Ernzerhof (PBE)  for exchange-correlation functions \citep{GGA}.
The electrons explicitly included in the calculations are the ($2s^22p^2$) electrons of carbon, 
the ($2s^22p^1$) electrons of boron and ($2s^22p^3$) electrons of nitrogen. 
The core electrons ($1s^2$) of carbon, boron and nitrogen are replaced by the projector augmented wave (PAW) and pseudopotential approach\citep{paw1}. 
A plane-wave cutoff of 520 eV is used in all the calculations.

The criterion to stop the relaxation of the electronic degrees of freedom is set by total energy change to be smaller than 0.000001 eV. 
The optimized atomic geometry was achieved through minimizing Hellmann-Feynman forces 
acting on each atom  until the maximum forces on the ions were smaller than 0.001 eV/\AA.

The atomic structures of the five configurations were obtained by fully relaxing a 24-atom-unit cell where all atoms were placed in one plane. 
The irreducible Brillouin Zone was sampled with a Gamma-centered $19 \times 19 \times 1$ $k$-mesh and initial charge densities were taken as a superposition of atomic charge densities. 
There was a 14 \AA \, thick vacuum region to reduce the inter-layer interaction to model the single layer system.

\section{RESULTS AND DISCUSSION}

\begin{figure}
\includegraphics[width=\ww\textwidth]{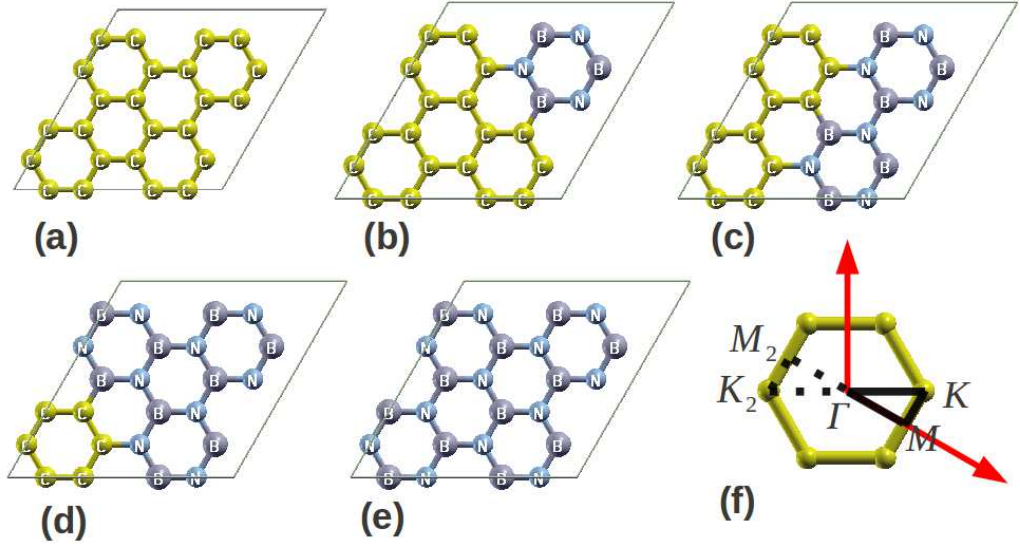}
\caption{\label{fig:config} Atomic structures of five configurations in order of h-BN concentration: (a) 0\%; (b) 25\%; (c) 50\%; (d) 75\%; (e) 100\%. (f) shows the Brillouin zone and high symmetrical $k$ points.} 
\end{figure}

    We examined the variation of band gap of h-BNC mono-layers as function of h-BN concentration. 
Five h-BNC configurations have been studied with 0\%, 25\%, 50\%, 75\% and 100\% concentration of h-BN, with 0\% and 100\% corresponding to pure graphene and h-BN, respectively. The three other concentrations were selected based on their simplicity and representativeness. 
The atomic structures of these five configurations (\fig{fig:config}) were determined by the {\em ab initio} density functional theory. 

The optimized atomistic structures of the five configurations shown in panels (a)-(e) in \fig{fig:config} 
were determined using DFT calculations. The lattice constants are listed in \tab{tab:a0}. 
Due to the intrinsic difference between pure h-BN and graphene, 
the lattice constants of the h-BNC mixtures are obtained by averaging the lattice vectors of the super-cells.
We found that the lattice constant increases with h-BN concentration $x$. 
Our results are in good agreement with experiments in h-BN (2.51 \AA) \citep{PhysRevB.68.104102}  
and graphene (2.46 \AA) \citep{ISI:A1955WB70900014}.  

\begin{table}
\center
\caption{\label{tab:a0} Lattice constant $a$, bottom edge of conduction bands $E_c$, top edge of valence bands $E_v$, band gap energy $E_g$
of the five configurations from DFT calculations.}
\begin{tabular}{|c|c|c|c|c|c|}
\hline
$x$ & 0.0 & 0.25 & 0.5 & 0.75 & 1.0 \\
\hline
$a$ (\AA) &2.468&2.484&2.496&2.506&2.512\\
\hline
$E_c$ (eV) &0&0.57&0.97&1.72&2.90\\
\hline
$E_v$ (eV)  &0&-0.44&-0.70&-1.15&-1.78 \\
\hline
$E_g$ (eV) &0&1.01&1.67&2.86&4.68\\
\hline
\end{tabular}
\end{table}

\begin{figure}
\vspace*{-10pt}
\includegraphics[width=\ww\textwidth]{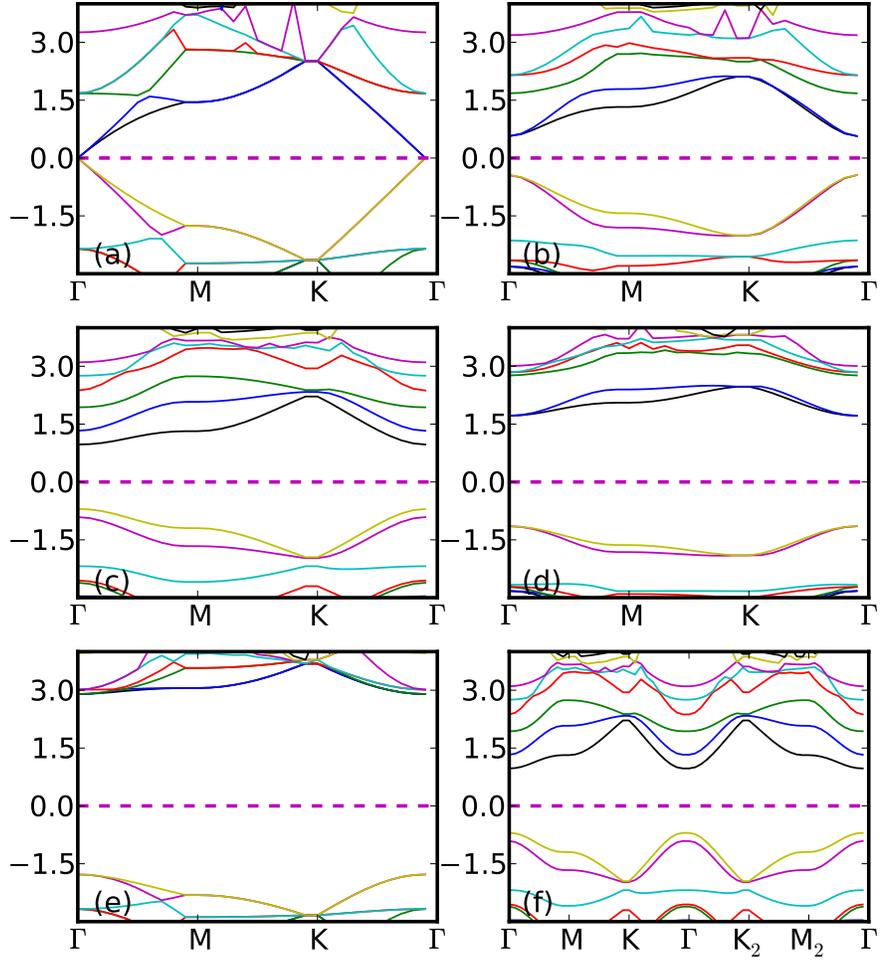}
\caption{\label{fig:band} Electronic band structures of five configurations in order of h-BN concentration: (a) 0.0; (b) 0.25; (c) 0.5; (d) 0.75; (e) 1.0. (f) shows symmetry of band structure in configuration (c).} 
\end{figure}

The electronic band structures of the optimized atomistic structures of the five configurations are plotted in \fig{fig:band} a-e, in order of $x$. 
The $\Gamma,M$ and $K$ are the high geometrically symmetrical $k$ points in the Brillouin zone, as noted in panel (f) in \fig{fig:config}. 
The dashed line in the band structure plots notes the Fermi energy of graphene, which is the intrinsic Fermi energy for h-BNC structures. 
The band energies are shifted so that the intrinsic Fermi surface energy is aligned up. 
The bottom edge of conduction bands $E_c$, top edge of valence bands $E_v$, and band gap energy $E_g$ of the five configurations from DFT calculations are also obtained and listed in \tab{tab:a0}. 
We notice that the band gap increases with $x$, with the curvature of the band at $\Gamma$ decreasing with $x$ indicating increment of effective mass. 
The symmetry of the real space is broken in h-BNC since the elements are heterogeneous. We tested the symmetry of the electronic band structure at the point of $M_2$ and $K_2$, which are center symmetrical points of $M$ and $K$ respectively. We found that the symmetry of electronic band structures at these points are still preserved, as shown in \fig{fig:band}f, where the band structure of configuration (c) is presented as an example.     

\begin{figure}[htp]
\includegraphics[width=\ww\textwidth]{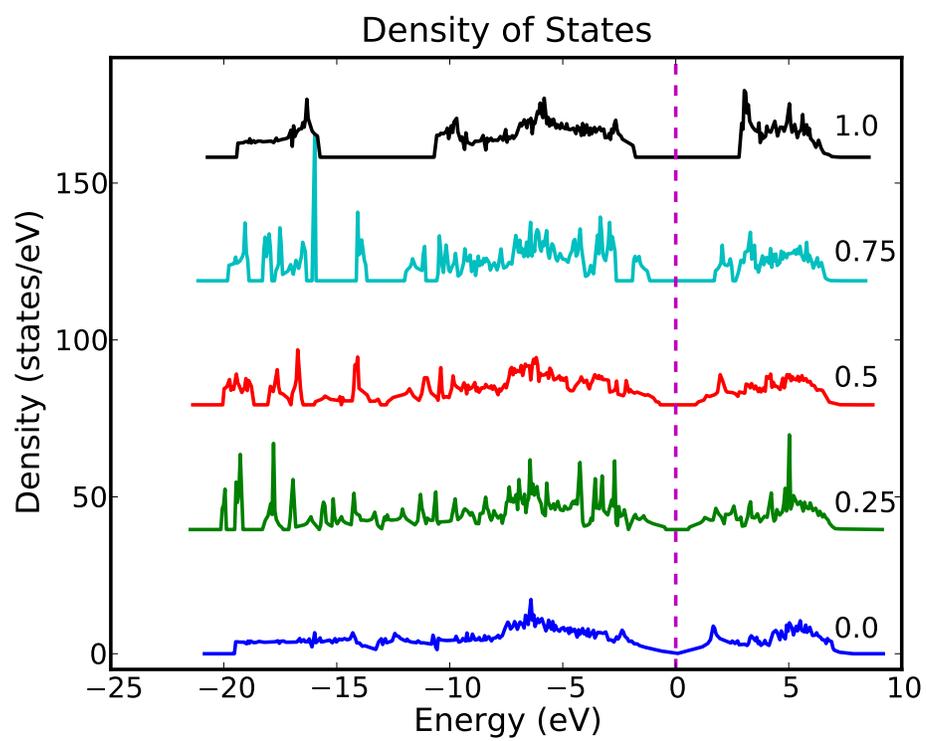}
\caption{\label{fig:dos} Density of states of five configurations in order of h-BN concentration.} 
\end{figure}

The density of states (DOS) of the optimized atomistic structures of the five configurations 
are plotted in \fig{fig:dos} in order of increasing h-BN concentration. 
The lines were shifted by 40 (states/eV) to avoid overlapping. 
The energies were shifted so that the Fermi surface energy could be set to zero. 
This DOS plot clearly shows the systematic increment of the band gap with respect to h-BN concentration.

\begin{figure}[htp]
\includegraphics[width=\ww\textwidth]{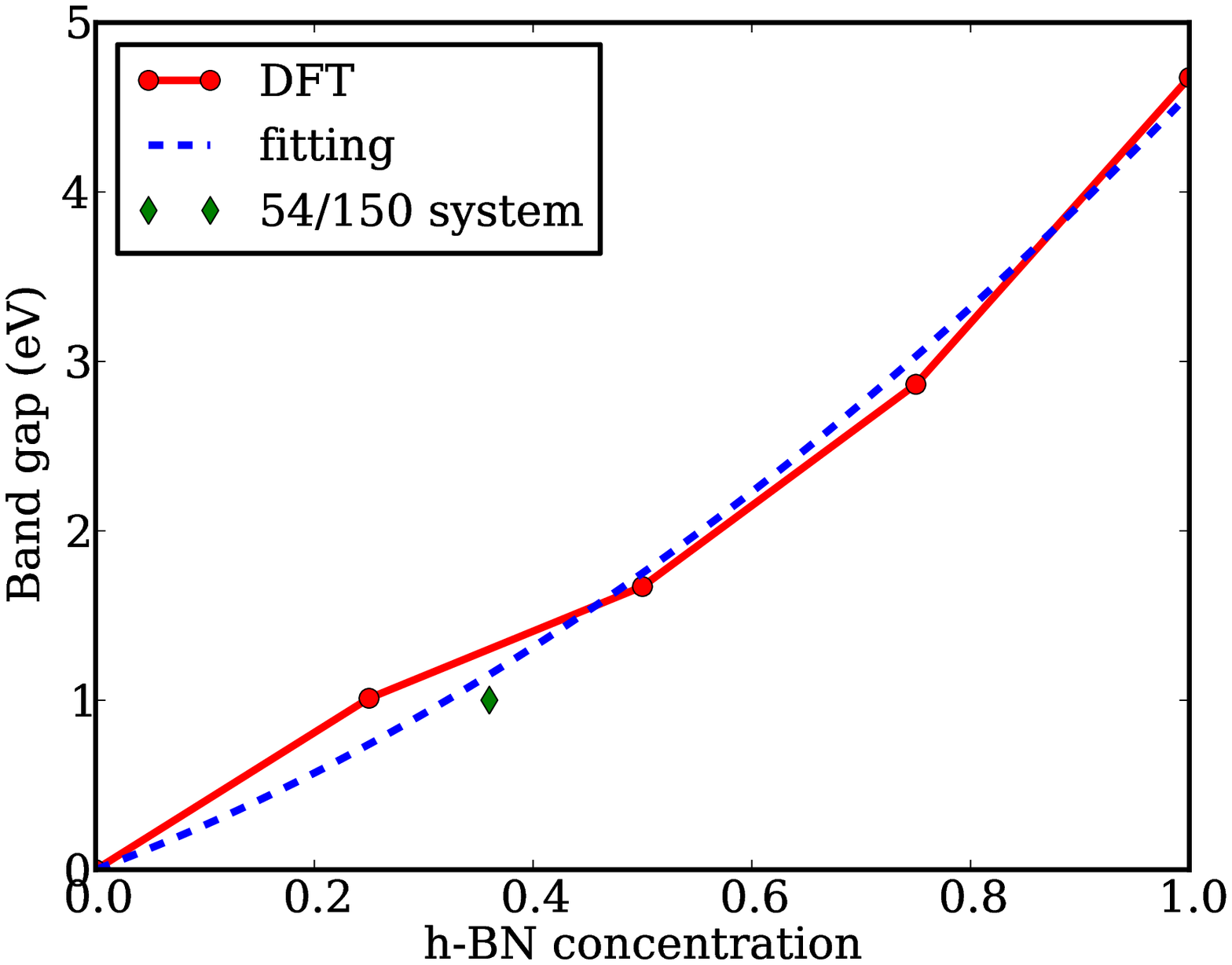}
\caption{\label{fig:Eg} Band gap energy $E_g$ as a function of h-BN concentrations $x$, fitted to a quadratic function. } 
\end{figure}

The band gap energy ($E_g$) of the optimized atomistic structures of the five configurations from DFT calculations are plotted in \fig{fig:Eg} 
as a function of $x$. Our calculation (DFT-GGA) of the $E_g$ of graphene-like h-BN (4.67 eV)  
is in good agreement with DFT-LDA prediction of 4.7 eV \citep{Giovannetti}.  
What is most interesting is that 
$E_g$ increases with $x$ in a nearly quadratic manner. 
The relationship between $E_g$ and h-BN concentrations may be approximated by $E_g(x)=2.16 x^2 + 2.42 x$ as plotted in dashed line in \fig{fig:Eg}.

\begin{figure}[htp]
\includegraphics[width=\ww\textwidth]{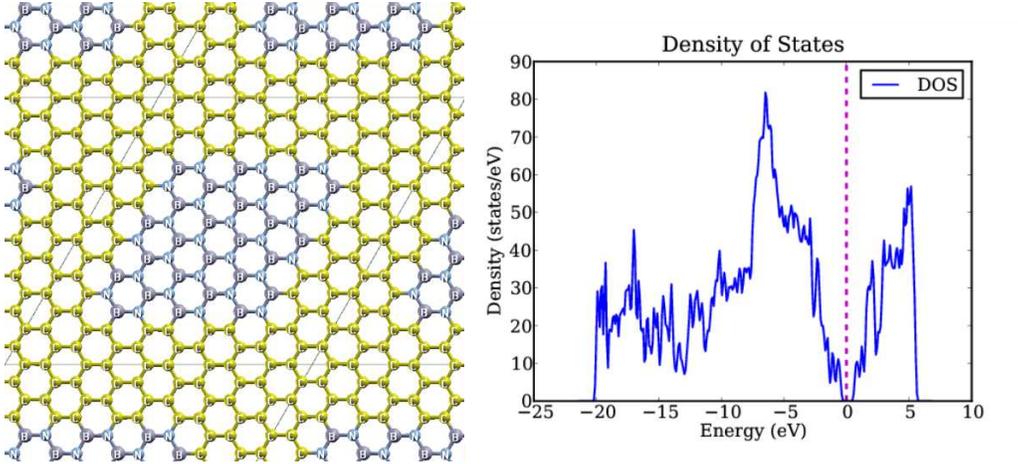}
\caption{\label{fig:big} The configuration (left) and Density of States (right) of the 54/150 (54 h-BN atoms in 150 h-BNC atoms) system.}
\end{figure}

The system size effect to the band gap is examined with a 150-atom system 
of which 54 atoms in h-BN phase form an ``island" surrounded by 96 atoms in graphene phase as shown in \fig{fig:big}.
The band gap energy of this 54/150 h-BNC system is 1.001 eV as the diamond dot in \fig{fig:Eg}, 
agreeing well with the prediction from 24-atom model. The full density of states is in plotted in \fig{fig:big}

A continuous and wide range of tunable band gap of h-BNC materials  is possible ranging from semi-metal (graphene) to semi-conductor ($E_g$ up to 4 eV) and insulator ($E_g>4$ eV). 
Interestingly, the band gap energies of h-BNC at $x=0.25,0.5,0.75$ 
(corresponding to BNC$_6$, BNC$_2$, B$_3$N$_3$C$_2$) 
are similar to those of Silicon (1.11 eV), Selenium (1.74 eV) and Silicon carbide (2.86 eV), respectively \citep{ssed}, which indicates potential substitutions for those semiconductors.

The mechanism of this band gap energy increment may be roughly explained with bond stretching caused by introducing h-BN in graphene. 
Since the lattice constant $a$ increases with the h-BN concentration, 
the average bond length increases, with a reduction in the overlap of the wave functions, 
which causes the band gap enlargement.    

\begin{figure}[htp]
\includegraphics[width=\ww\textwidth]{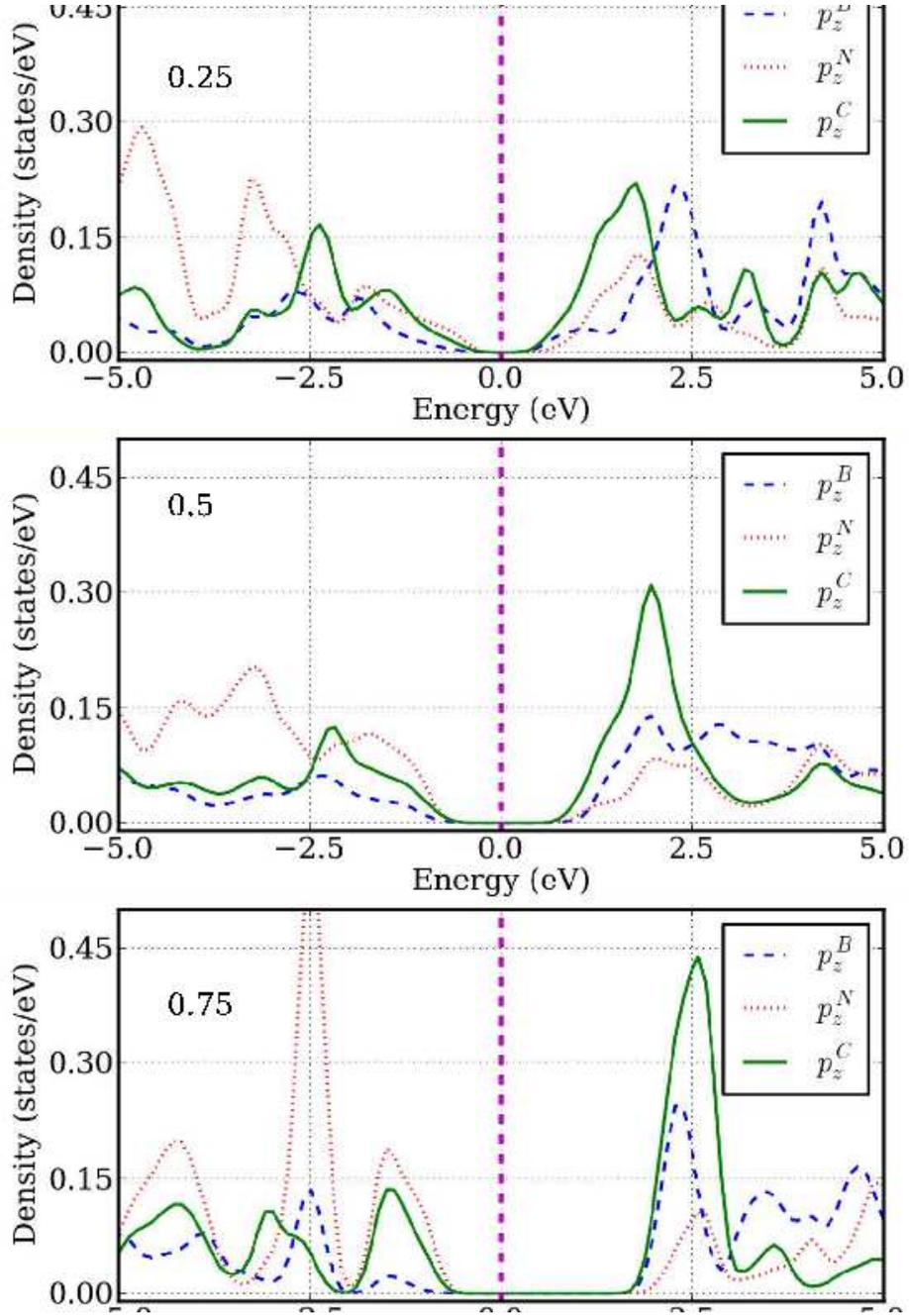}
\caption{\label{fig:pz} Projected density of states of the atoms in the interface of h-BN and graphene.} 
\end{figure}

We also studied the projected density of states (PDOS) of the atoms in the system. 
Each atom in the system has the same band gap in PDOS as that of total system.
The the atoms on the interface of h-BN and graphene were plotted in \fig{fig:pz}. 
Only $P_z$ electrons, which form $\pi$ bonds in $z$ direction, contribute to the band gap. 
The edge of the band gap becomes sharper when the concentration of h-BN is higher, indicating the higher localization of electrons around the Fermi surface. 

\begin{figure}[htp]
\includegraphics[width=\ww\textwidth]{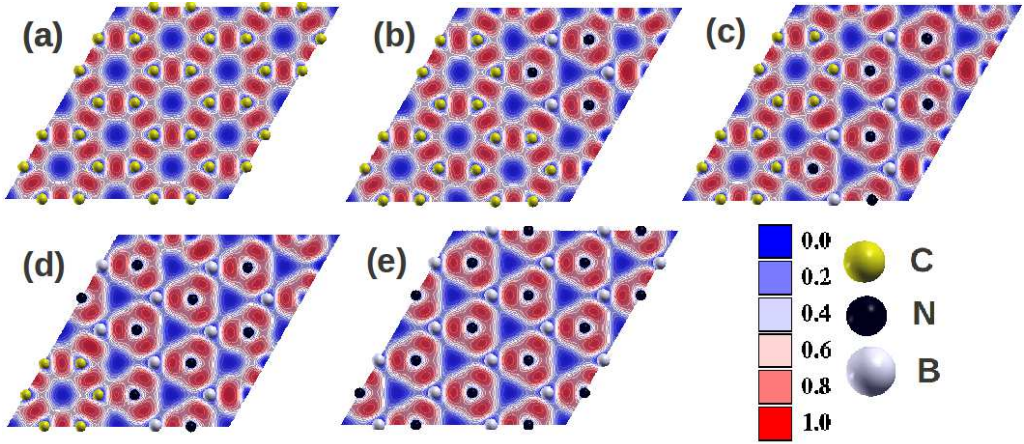}
\caption{\label{fig:elf} Electron localization function in the five configurations.} 
\end{figure}

In particular, since the total number of electrons of a h-BNC system is the same as in a corresponding graphene structure, the mechanism of this band gap tuning may be ascribed to electron redistribution, bonding and localization due to many-body interactions in the system. 
The ``electron localization function" (ELF) measures the extent of spatial localization of the reference electron \citep{BECKE1990}, 
with values between 0 and 1. Values of 0, 0.5 and 1.0 correspond to vacuum, uniform electron gas and perfect localization, respectively. 
A higher ELF value at a point indicates that the electrons are more localized there than in a uniform electron gas of the same density. 
ELF plots of the five configurations from DFT calculations are shown in 
\fig{fig:elf} a-e.  
The plots show that electrons are more localized around nitrogen atoms, 
which is consistent with the fact that nitrogen atoms have more valence electrons than carbon and boron atoms. 
The electrons on the B-N and B-C bonds are more localized than the C-C and N-C bonds. 
Since the increment of h-BN concentration enhances overall electron localization, the band gap energy enlarges as a consequence. 

\section{CONCLUSIONS}

    In summary, we used {\em ab initio} density functional theory 
to investigate the effect of the h-BN domain size on 
the band gap of h-BNC hybrid structures with 
(B$_3$N$_3$)$_x$(C$_6$)$_{1-x}$ model.  
This simple model of h-BNC heterostructure with phase segregation 
not only considers the stoichiometry, but also the shape of the nano-structure: 
which should be hexagonal ring of (B$_3$N$_3$) or (C$_6$).
The atomic structures, electronic band structures, 
density of states, projected density of states, electron localization functions 
of five h-BNC hybrid structures were explicitly examined. 
We found that the band gap energy of h-BNC can be continuously 
tuned in full range between that of two phases,
graphene and h-BN, as a function of h-BN concentration.
The origin of the tunable band gap in these heterosturcutes
are due to the change in electron localization 
by changing the h-BN concentration.
This result may provide a new approach in practical 
engineering applications of these heterostructures. 

\section*{ACKNOWLEDGEMENTS}

The authors would like to acknowledge the generous financial support from the Defense Threat Reduction Agency (DTRA) Grant \# BRBAA08-C-2-0130.

\bibliographystyle{elsarticle-num}

\bibliography{bnc}

\end{document}